\documentclass[epj]{svjour}
\usepackage{graphics}
\begin{document}
\title{Hybrid Stars in a Strong Magnetic Field}
\author{V. Dexheimer\inst{1,2} \and R. Negreiros\inst{3,4} \and S.
Schramm\inst{3}%
}                     
\offprints{}          
\institute{UFSC, Florianopolis, Brazil \and Gettysburg College, Gettusburg, PA,
USA \and FIAS, Johann Wolfgang Goethe University, Frankfurt, DE \and Instituto de Fisica, UFF, Av. Gal. Milton Tavares de Souza s/nº. Gragoata, Niteroi, 24210-346, Brazil}

\date{Received: date / Revised version: date}
%
\abstract{We study the effects of high magnetic fields on the particle
population and equation of state of hybrid stars using an extended hadronic and
quark SU(3) non-linear realization of the sigma model. In this model the degrees
of freedom change naturally from hadrons to quarks as the density and/or
temperature increases. The effects of high magnetic fields and anomalous
magnetic moment are visible in the macroscopic properties of the star, such as
mass, adiabatic index, moment of inertia, and cooling curves. Moreover, at the same time that the magnetic fields become high enough to modify those properties, they make the star anisotropic.
} 
\maketitle
\section{Introduction}
\label{intro}

Magnetars are compact stars that have surface magnetic fields up to
$10^{14}-10^{15}$ G \cite{Paczynski:1992zz,Ibrahim:2003ev}. Such fields can be
estimated from observations of the star's period and period derivative.
According to Virial theorem estimates, neutron stars could have a central
magnetic field as large as $10^{18}$ or $10^{19}$ G. An
accurate calculation of this limit is complicated, since all energies to be
weighed against the magnetic energy (the one from matter and the gravitational
one) also depend on the magnetic field (due to the nonlinear nature of General
Relativity). For this reason the limit might be different for different
equations of state (EOS's). Furthermore, due to the asymmetry introduced by the magnetic
field in the z-direction, the pressure becomes different in the parallel and
perpendicular directions. In reality, depending on the magnitude of the magnetic
field, the parallel pressure becomes much smaller than the perpendicular one,
and in extreme cases, can go locally to zero. In Ref.~\cite{Ferrer:2010wz}, as
well as in our 
calculations, this limit was found to be around $10^{18}$ - $10^{19}$ G. Beyond
this value strong instabilities can occur, which indicates that further
investigation is necessary.

References \cite{Bocquet:1995je,2001ApJ...554..322C} solve the Einstein and
Maxwell equations self-consistently for non-rotating and rotating stars. They
study axisymmetric and poloidal magnetic field configurations and find
$B\sim0.1$ to $4.2\times10^{18}$ G as limits for the star central magnetic
field. This range arises from the use of different hadronic EOS's,
from a simple politropic to a hyperonic relativistic one. Although such
EOS do not take into account magnetic field modifications,
 their formalism provides reliable results for star masses and radii because it
takes into consideration different pressures in different directions of the
star. It is important to notice that Ref.~\cite{Broderick:2001qw} states that
the use of different current functions and symmetries, other than the axial
one, together with effects of the magnetic field on the EOS may
alter these limits. Other simulations including magnetic field effects can be
found in \cite{1954ApJ...119..407F,Lander:2012iz,Rezzolla:1999he,Kiuchi:2007pa}.

It has also been shown by
refs.~\cite{2010PhRvD..81j5021K,Kurkela:2010yk} that using well accepted
hadronic EOS's together with those based on first principles QCD calculations,
one finds a phase transition to deconfined matter already at a few times
saturation density inside compact stars. Such a phase transition can be a sharp
first order transition or might exhibit a mixed phase, depending on the
surface tension between the phases. In this work we study the effect of the
latter. 

There have also been works
\cite{PhysRevLett.97.122001,Baym:2008me,2012APS..APRD10001P,Bratovic:2012qs,%
Lourenco:2012dx,Lourenco:2012yv,Yamamoto:2007ah,Yamamoto:2008zw,Abuki:2010jq}
showing that the inclusion of sophisticated quark coupling terms (like vector
couplings, effective 6-quark interactions and quark-quark pairing) to the NJL
and PNJL models, can lead to the appearance of a crossover chiral symmetry
restoration/deconfinement phase transition at high density and small
temperatures. In this case, the QCD phase diagram becomes more complicated due
to the presence of a second critical point. Unfortunately, the values of the
coupling constants which determine the precise nature of the phase transition in
this regime are not predictable from first principles QCD calculations at this
point.  Thus, the determination of the nature of the phase
transition in this limit is still an active area of research.

One important question we will address is whether the effects of strong
magnetic fields in the deconfinement phase transition are strong enough to be
observed. Other studies along this line can be found in
\cite{Chakrabarty:1996te,Bandyopadhyay:1997kh,Rabhi:2009ih,Paulucci:2010uj}. In
order to do so in a realistic way, we adopt as our microscopic model a chiral
approach that includes hadronic as well as quark degrees of freedom in a unified
description \cite{Dexheimer:2009hi,Schramm:2011aa,Negreiros:2010tf}. Here,
besides the analysis of the mass-radius diagram, we will extend
our study to the adiabatic index and moment of inertia of the stars.
Furthermore, we will, for the first time, investigate the thermal evolution of
highly magnetized hybrid stars that include modifications in the EOS, which will
help us to assess the effect of a high magnetic field on the cooling of the
star. Finally, we will use two different approximations to take into account the pure magnetic field contribution to the energy density and pressure of the system. With this, we can model an EOS that is in principle anisotropic in an isotropic way and try to determine how reliable those approximations are.

\section{The Model}
\label{the model}

Chiral sigma models are effective quantum relativistic models that describe
hadrons interacting via meson exchange and, most importantly, are constructed
from symmetry relations. They are constructed in a chirally invariant manner as
the particle masses originate from interactions with the medium and, therefore,
go to zero at high density and/or temperature. Adopting the nonlinear
realization of the sigma model gives a significant improvement to the widely
used linear sigma model \cite{Papazoglou:1997uw,Lenaghan:2000ey} as it is in
better agreement with nuclear physics results
\cite{Papazoglou:1998vr,Bonanno:2008tt}. 

The Lagrangian density of the SU(3) non-linear sigma model in the mean field
approximation constrained further by astrophysics data can be found in
Ref.~\cite{Dexheimer:2008ax}. A recent extension of this model also includes
quarks as dynamical degrees of freedom \cite{Dexheimer:2009hi}. In this version,
the degrees of freedom change due to the effective masses of the baryons and
quarks. Their masses are generated by the scalar mesons ($\sigma$, isovector
$\delta$, strange $\zeta$), except for a small explicit mass term $M_0$ and the
term containing the field $\Phi$. This field serves as an effective order
parameter for the deconfinement transition and is modeled in such a way as to
reproduce the behavior of the Polyakov loop at zero chemical potential, as
determined by lattice QCD calculations \cite{Fukushima:2003fw}. The baryon and
quark effective masses are given by
\begin{equation}
M_{B}^*=g_{B\sigma}\sigma+g_{B\delta}\tau_3\delta+g_{B\zeta}\zeta+M_{0_B}+g_{
B\Phi} \Phi^2,
\label{6}
\end{equation}
\begin{equation}
M_{q}^*=g_{q\sigma}\sigma+g_{q\delta}\tau_3\delta+g_{q\zeta}\zeta+M_{0_q}+g_{
q\Phi}(1-\Phi),
\label{7}
\end{equation}
where the values for the coupling constants can be found in
Ref.~\cite{Dexheimer:2009hi}. With the increase of density/temperature, the
$\sigma$ field (non-strange chiral condensate) decreases its value, causing the
effective masses of the particles (in the absence of $\Phi$) to decrease towards chiral symmetry
restoration. The field $\Phi$ assumes non-zero values with the increase of
temperature/density and, due to its presence in the baryons effective mass
(Eq.~(\ref{6})), suppresses their presence. On the other hand, the presence of
the $\Phi$ field in the effective mass of the quarks, included with a negative
sign (Eq.~(\ref{7})), ensures that they will not be present at low
temperatures/densities. The value for the $g_\Phi$ coupling constants is
intrinsically related to the strength of the phase transition and only high
values reproduce a first order phase transition at high densities and low
temperatures.

\begin{figure}
\resizebox{0.48\textwidth}{!}{%
\includegraphics{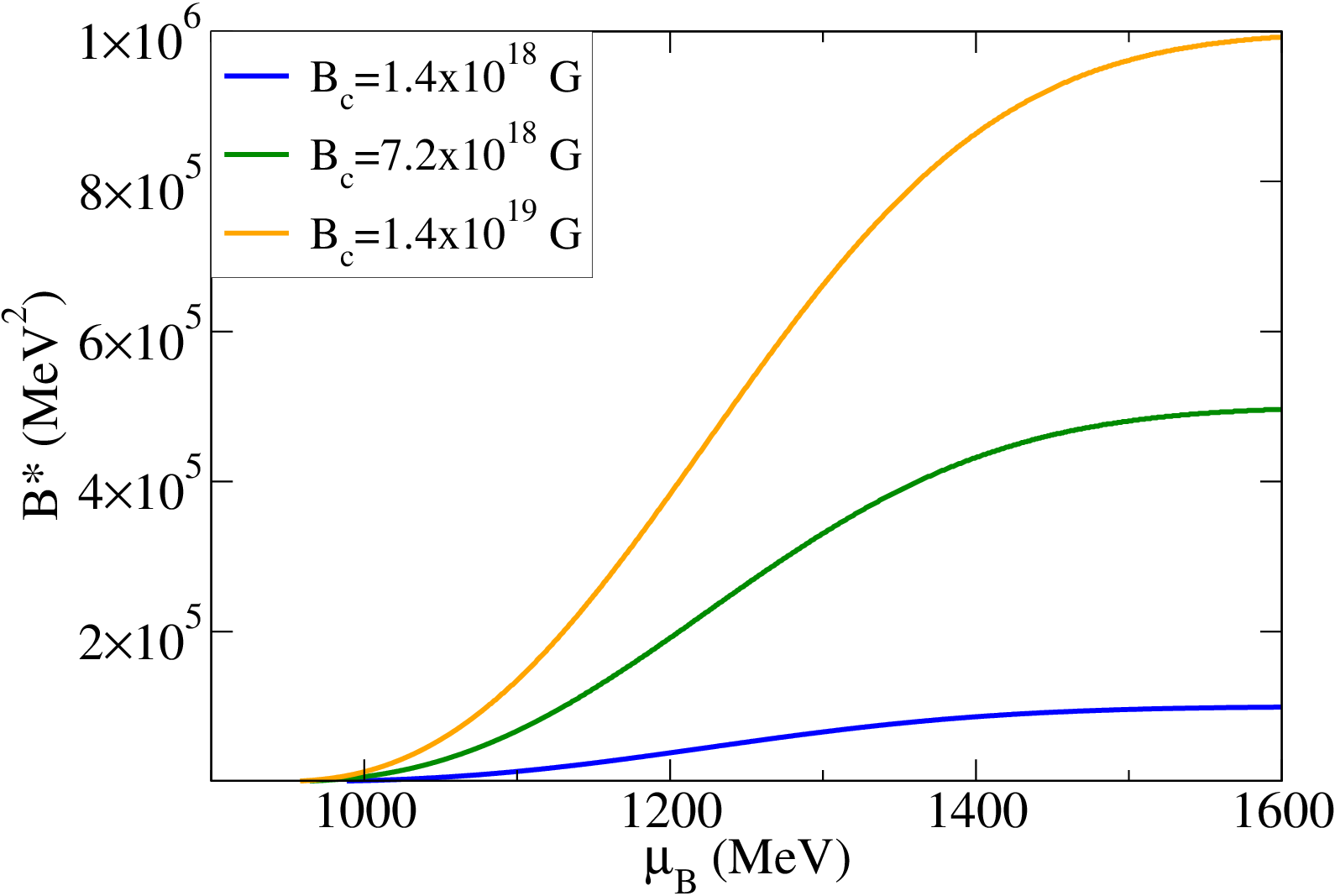}}
\caption{(Color online) Effective magnetic field as a function of baryon
chemical potential shown for different central magnetic fields.}
\label{Beff}
\end{figure}

Both phase transitions, chiral symmetry restoration (breaking) and deconfinement
(confinement) happen at the same density/temperature. Such a fact comes from the
correlation of these quantities in the effective masses of the particles. The
potential for $\Phi$ reads
\begin{eqnarray}
U=(a_0T^4+a_1\mu_B^4+a_2T^2\mu_B^2)\Phi^2\\ \nonumber
+a_3T_0^4\log{(1-6\Phi^2+8\Phi^3-3\Phi^4)}.
\end{eqnarray}
This potential was modified from its original form in the PNJL model
\cite{Ratti:2005jh,Roessner:2006xn} by adding terms that depend on the baryon
chemical potential. This allows us to reproduce the phase structure over the
whole range of chemical potentials and temperatures, as suggested in lattice QCD
studies \cite{Fodor:2004nz,Aoki:2006we}, i.e. a cross over at small chemical potential and,
at higher chemical potential, a first-order transition line that stops at a
second-order critical end-point. Here, by using the modified $\mu_B$
dependent form for the $\Phi$ potential $U$, combined with a high value for the
coupling of the $\Phi$ field for the hadrons and quarks, we obtain a first order
phase transition at high densities, as it has been conjectured in other hybrid
star calculations
\cite{Lin:2010zw,Lenzi:2010mz,Agrawal:2010er,Chen:2011my,Endo:2011em}. We stress
again that this behavior is only a model assumption and, for instance in a
Polyakov-extended quark-meson model \cite{Herbst:2010rf} or in the 
quarkyonic picture \cite{McLerran:2007qj} one might expect a smooth
deconfinement crossover in this regime.

\section{Inclusion of Magnetic Field}
\label{inclusion of magnetic field}

In order to have a more complete description of hybrid stars, we include
magnetic fields that, while being in the z-direction, are not constant. The
effective magnetic dipole field $B^*$ increases with baryon chemical potential
going from a surface value of $B_{surf}=10^{15}$ G (when
$\mu_B\simeq 938$ MeV) to a central value $B_c$ at high baryon
chemical potential
\begin{equation}
B^*(\mu_B)=B_{surf}+B_c[1-e^{b\frac{(\mu_B-938)^a}{938}}],
\end{equation}
with $a=2.5$, $b=-4.08\times10^{-4}$ and $\mu_B$ given in MeV. The use of the
formula above generates no discontinuity in the effective magnetic field or in
the increase of the effective magnetic field at the phase transition. Such an
unphysical discontinuity would be present if we had chosen the effective
magnetic field to be a function of baryon density, as is normally the case.
The constants $a$, $b$ and the form of $B^*$ are chosen to reproduce (in the absence
of quarks) the effective magnetic field curve as a function of density from
Refs.~\cite{Bandyopadhyay:1997kh,Mao:2001fq,Rabhi:2009ih}. As can be seen in
Fig.~\ref{Beff}, even with the use of extremely high central magnetic fields, the
values for the effective magnetic fields only become extreme at very high
baryon chemical potentials. In practice, only about $70\%$ of $B_c$ can be
reached inside the star. Still, the highest central magnetic field considered in
this work is above 
the limit established by hydrostatic stability from
refs.~\cite{Bocquet:1995je,2001ApJ...554..322C} and is only shown to illustrate
the influence of an extreme high magnetic field in the EOS.

\begin{figure}
\resizebox{0.48\textwidth}{!}{%
\includegraphics{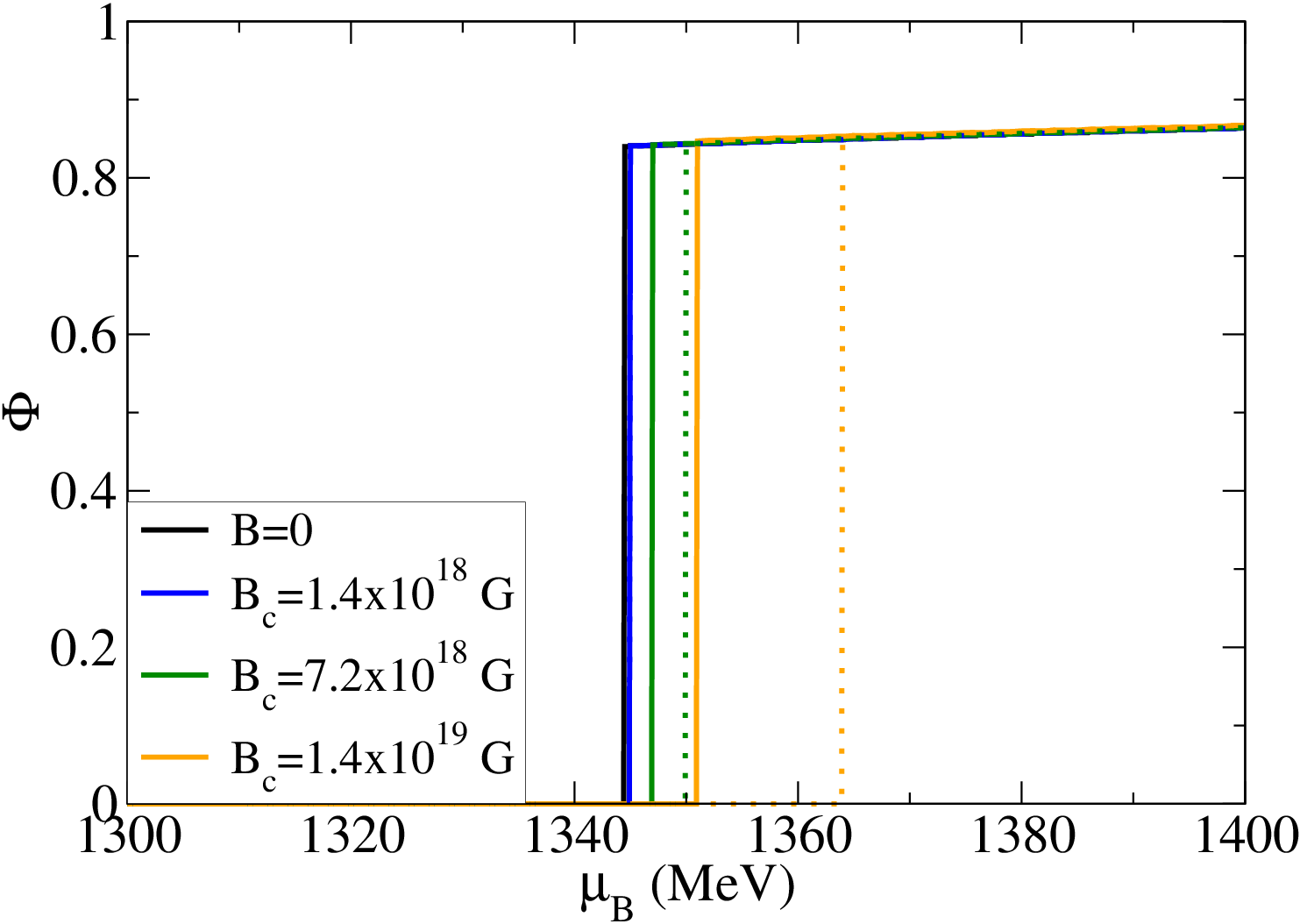}}
\caption{(Color online) Deconfinement order parameter as a function of baryon chemical
potential imposing local charge neutrality shown for different central magnetic
fields. Dotted lines include AMM.}
\label{pol}
\end{figure}

The magnetic field in the z-direction forces the eigenstates in the x and y
directions of the charged particles to be quantized into Landau levels $\nu$
\begin{equation}
E_{i_{\nu s}}^*=\sqrt{k_{z_{i}}^2+\left(\sqrt{m_i^{*2}+2
\nu|q_i|B^*}-s_i\kappa_i B^* \right)^2},
\end{equation}
where $k_i$ is the fermi momentum, $q_i$ the charge and $s_i$ the spin of each
baryon or quark. The last term comes from the anomalous magnetic moment (AMM) of
the particle that splits the energy levels with respect to the
alignment/anti-alignment of the spin with the magnetic field. The AMM also
modifies the energy levels of the uncharged particles
\begin{equation}
E_{i_{s}}^*=\sqrt{k_{i}^2+\left({{m_i^*}^2}-s_i\kappa_i B^* \right)^2}.
\end{equation}
The constants $\kappa_i$ determine the tensorial coupling strength of baryons
with the electromagnetic field tensor and have values $\kappa_p=1.79$,
$\kappa_n=-1.91$, $\kappa_\Lambda=-0.61$, $\kappa_\Sigma^+=1.67$,
$\kappa_\Sigma^0=1.61$, $\kappa_\Sigma^-=-0.38$, $\kappa_\Xi^0=-1.25$,
$\kappa_\Xi^-=0.06$. The sign of $\kappa_i$  determines the preferred
orientation of the spin with the magnetic field. The sum over the Landau levels
$\nu$ runs up to a maximum value, beyond which the momentum of the particles in
the z-direction would be imaginary
\begin{equation}
\nu_{max} =\frac{{E_{i_s}^*}^2+s_i\kappa_i B^* - {m_i^*}^2 }{2|q_i|B^*}.
\end{equation}

We choose to include in our calculations the AMM effect for the hadrons only,
since the coupling strength of the particles $\kappa_i$ depends on the
corresponding magnetic moment, that up to now is not fully understood for the
quarks. Furthermore, it is stated in Ref.~\cite{Weinberg:1990xm}, that quarks in
the constituent quark model have no anomalous magnetic moment. For calculations
including AMM effects for the quarks see
refs.~\cite{Chakrabarty:1996te,Suh:2001tr,PerezMartinez:2005av,Felipe:2007vb}.
The AMM for the electrons is also not taken into account as its effect is
negligibly small. Note that the AMM removes the spin degeneracy of the particles
and their energy levels are further split in two levels each, which increases
even more the particle chemical potentials.

\begin{figure}
\resizebox{0.49\textwidth}{!}{%
\includegraphics{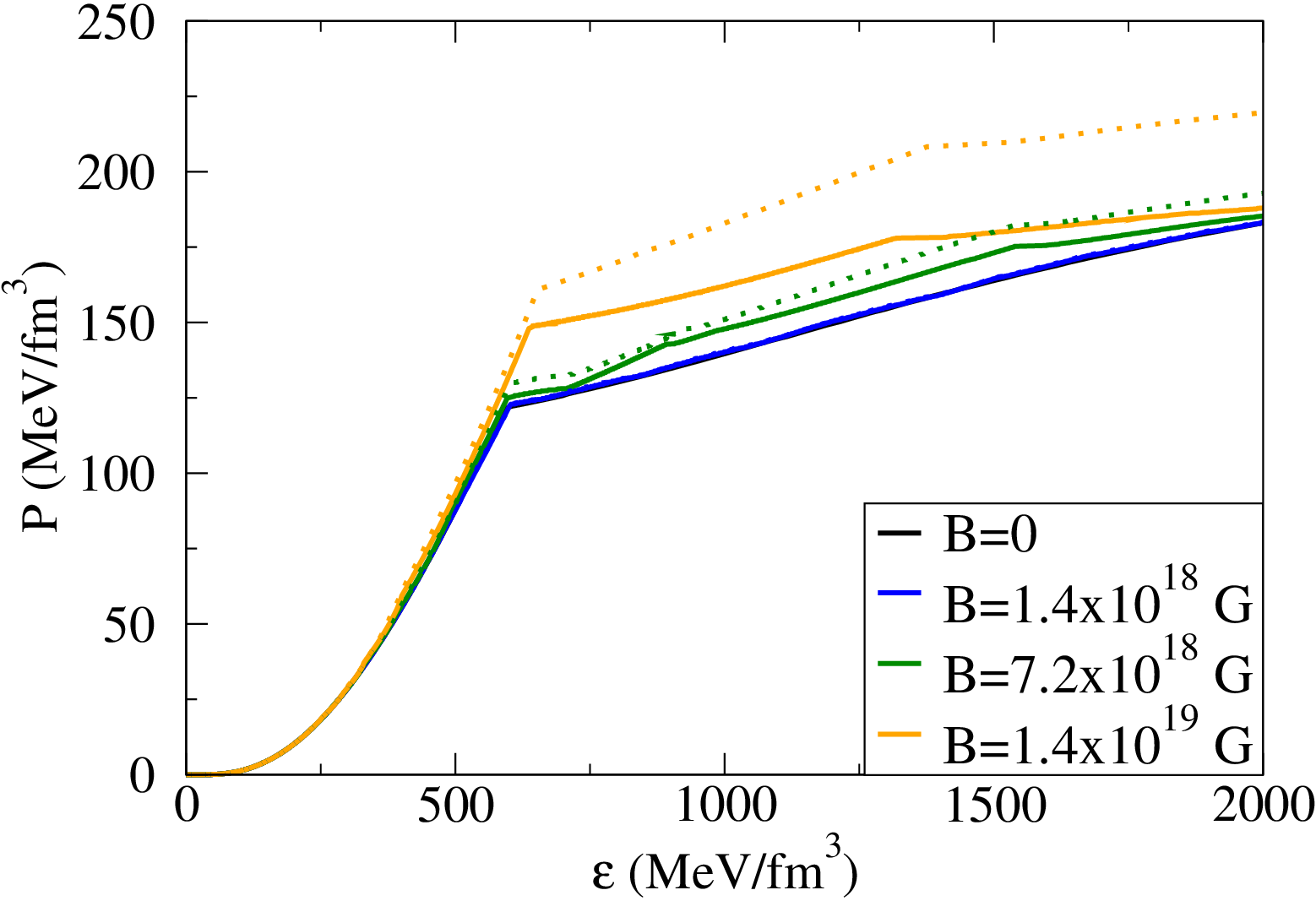}}
\caption{(Color online) Equation of State (pressure of matter as a function of
energy density of matter) assuming global charge neutrality for different
central magnetic fields. Dotted lines include AMM.}
\label{EOS}
\end{figure}

As can be seen in Fig.~\ref{pol}, a phase transition which is of first order at
zero temperature is delayed with the inclusion of high magnetic field.
This effect is due to a stronger stiffening of the hadronic part of the EOS (as
the magnetic field increases). The delay of the phase transition is further increased when the anomalous
magnetic moment is considered, as it only affects the hadronic phase, rendering
its EOS stiffer (due an increase in the chemical potential of the baryons). If
global charge neutrality is assumed, the location of the mixed phase does not
change substantially, except for the highest magnetic field considered with AMM,
when the entire mixed phase is pushed to slightly higher chemical potentials.

The same effect of confinement and/or chiral symmetry enhancement in the
presence of high magnetic fields was already predicted by other works. In
refs.~\cite{2009PhRvC..79c5807M,2009PhRvC..80f5805M,2011PhRvC..83f5805A} this
was shown for high density and small or zero temperature, whereas in
refs.~\cite{2010PhRvD..82e1501D,2010PhRvD..82j5016M,2011PhRvC..83f5805A,%
Gatto:2010pt} this was shown for small or zero density and high
temperatures. Together, these features show the importance of the study of the
influence of high magnetic fields and deconfinement/chiral symmetry restoration
in compact stars as a part of a greater picture that forms the whole QCD phase
diagram and contains heavy-ion experiments in the other extreme.

The EOS is shown in Fig.~\ref{EOS} for the case in which global
charge neutrality is allowed and a mixed phase appears. We do not show the pure
quark phase region, since in this model it is present only at very high
densities that cannot be reached in the interior of stars. It is important to notice that at very low
densities, the hadronic EOS of non-interacting matter gets softer in the
presence of high magnetic field, in agreement with
refs.~\cite{Broderick:2000pe,Broderick:2001qw} and other papers. In our
case this effect is smaller because the magnetic field is only large at high
densities. Note that only the energy density and pressure of matter are shown in this figure and no direct contribution from the magnetic field is included. In this way we can see the direct effects of the Landau quantization and the AMM in the model.

\begin{figure}
\resizebox{0.48\textwidth}{!}{%
\includegraphics{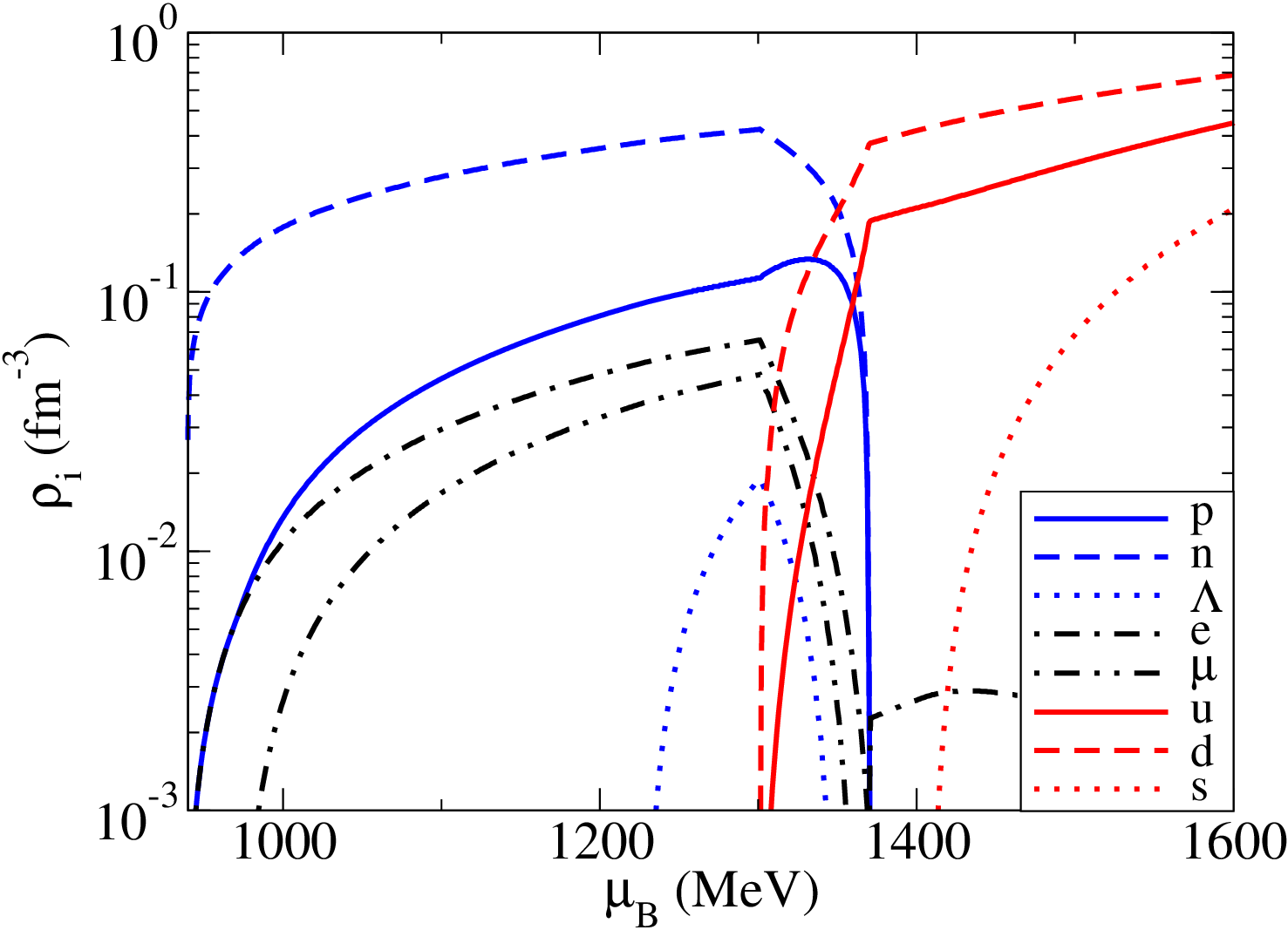}}
\caption{(Color online) Particle densities as a function of baryon chemical
potential assuming global charge neutrality for B=0.  Quark densities are
divided by $3$.}
\label{pop}
\end{figure}

Fig.~\ref{pop} shows the baryon density of fermions. In the mixed phase the
hadrons disappear as the quarks smoothly appear. The hyperons, despite being
included in the calculation, are suppressed by the appearance of the quark
phase. Only a very small amount of $\Lambda$ baryons appear right before the
phase transition. The density of electrons and muons is significant in the
hadronic phase but not in the quark phase. The reason for this behavior is that,
because the down and strange quarks are also negatively charged, there is no
necessity for the presence of electrons to maintain charge neutrality, and only
a small amount of leptons remains to ensure beta equilibrium. The strange quarks
appear after the other quarks, and do not make substantial changes in the
system. 

\begin{figure}
\resizebox{0.48\textwidth}{!}{%
\includegraphics{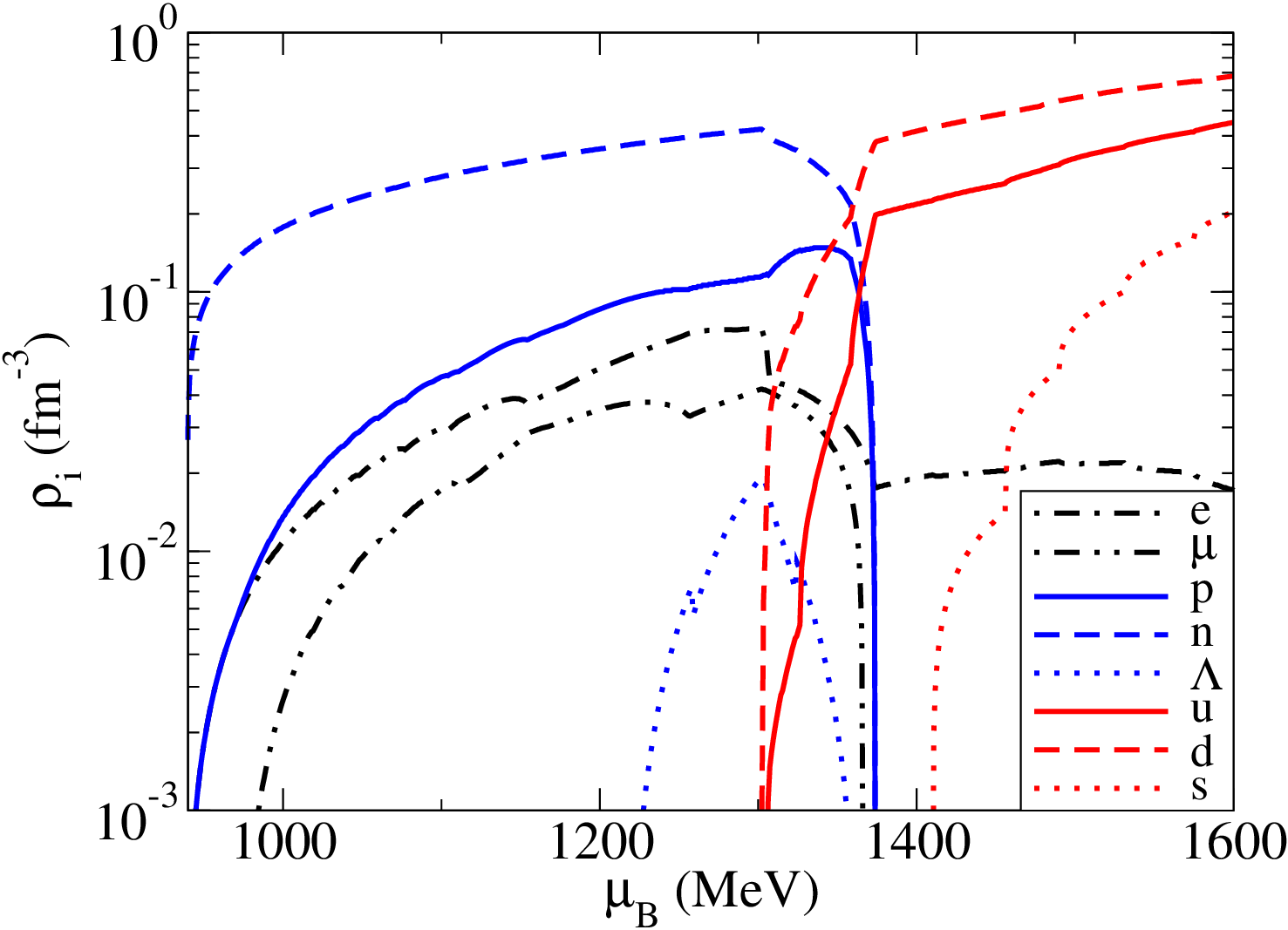}}
\caption{(Color online) Particle densities as a function of baryon chemical
potential assuming global charge neutrality for $B_c=7.2\times10^{18}$ G, and 
including AMM. Quark densities are divided by $3$.}
\label{popB}
\end{figure}

Fig.~\ref{popB} shows the change in population when a central magnetic field of
$7.2\times10^{18}$ G with AMM is considered. The wiggles in the charge particle
densities come from the Landau levels, more precisely when the Fermi energy of
the particles crosses the discrete threshold of a Landau level. The charged
particles are enhanced (as their chemical potentials increase with $B$), which
is especially visible by the amount of electrons in the quark phase. The
remaining hyperons are even more suppressed due to the increase in the proton
density. All these effects are further enhanced when higher central magnetic
fields are considered, like $B_c=1.4\times10^{19}$ G. 

\section{Macroscopic Properties}
\label{macroscopic properties}

A key point of this investigation is to determine whether the changes due to the
presence of magnetic field in the EOS are strong enough to affect observable
properties of the stars. To answer this question, we are going to analyze the
adiabatic index, moment of inertia, mass-radius diagram and thermal evolution
for different central magnetic fields. 

\subsection{Adiabatic Index}
\label{adiabatic index}

\begin{figure}
\resizebox{0.44\textwidth}{!}{%
\includegraphics{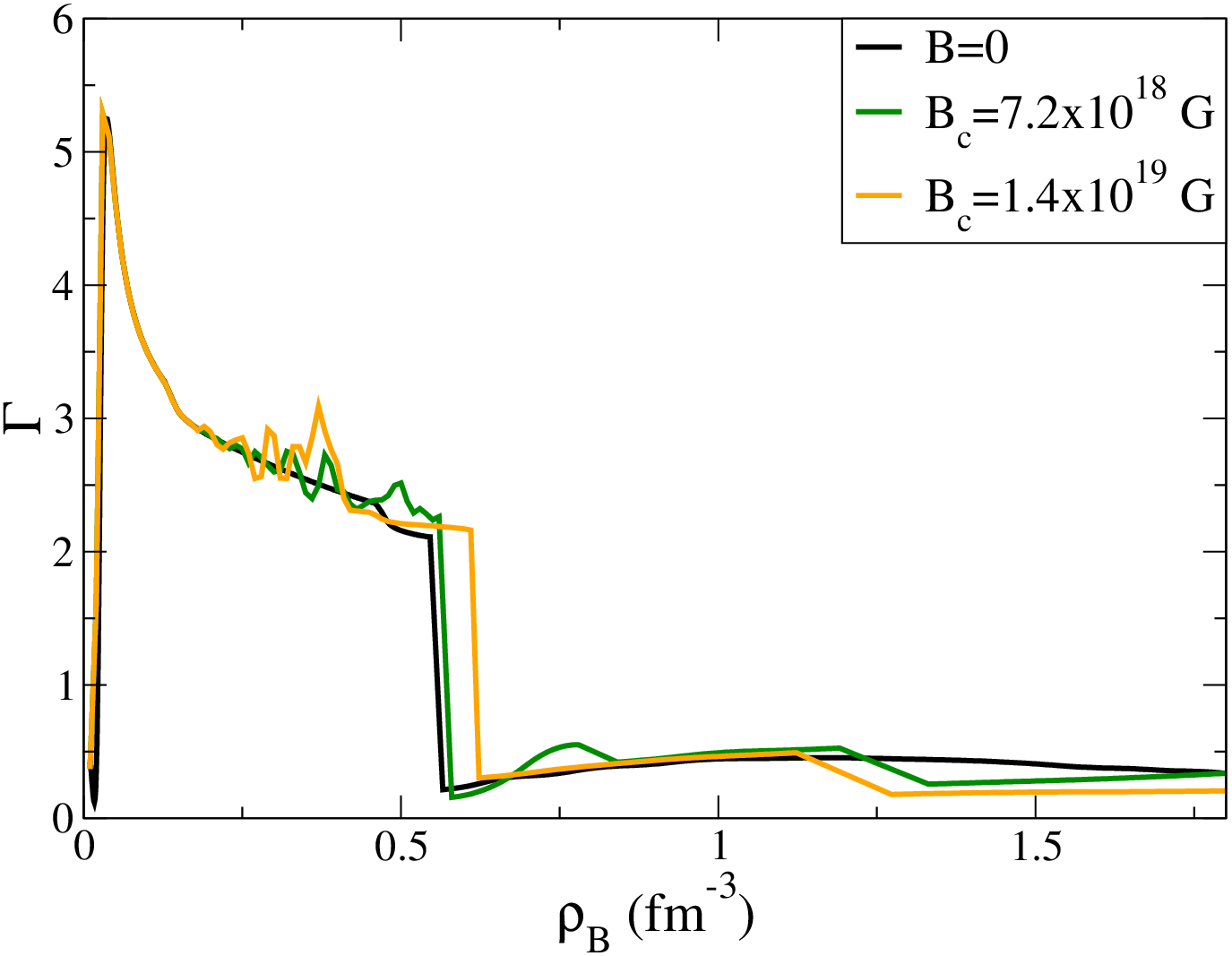}}
\caption{(Color online) Adiabatic index as a function of baryon density for
different central magnetic fields including AMM.}
\label{adiab}
\end{figure}

We begin with the adiabatic index (Fig.~ \ref{adiab}). The discontinuities in
the $B=0$ curve show the appearance of the $\Lambda$'s and the beginning of the
mixed phase. The extra peaks in the finite magnetic field with AMM curves show
the Landau level thresholds that also can be seen in the population plot. Note
that the Landau level thresholds appear in the hadronic phase as well as in the
mixed phase. As pointed out in Ref.~\cite{Ryu:2011gq}, these rapid changes in
the pressure can cause instabilities in the star that might cause star-quakes,
glitches and giant flares.

\subsection{Moment of Inertia}
\label{moment of inertia}

\begin{figure}
\resizebox{0.48\textwidth}{!}{%
\includegraphics{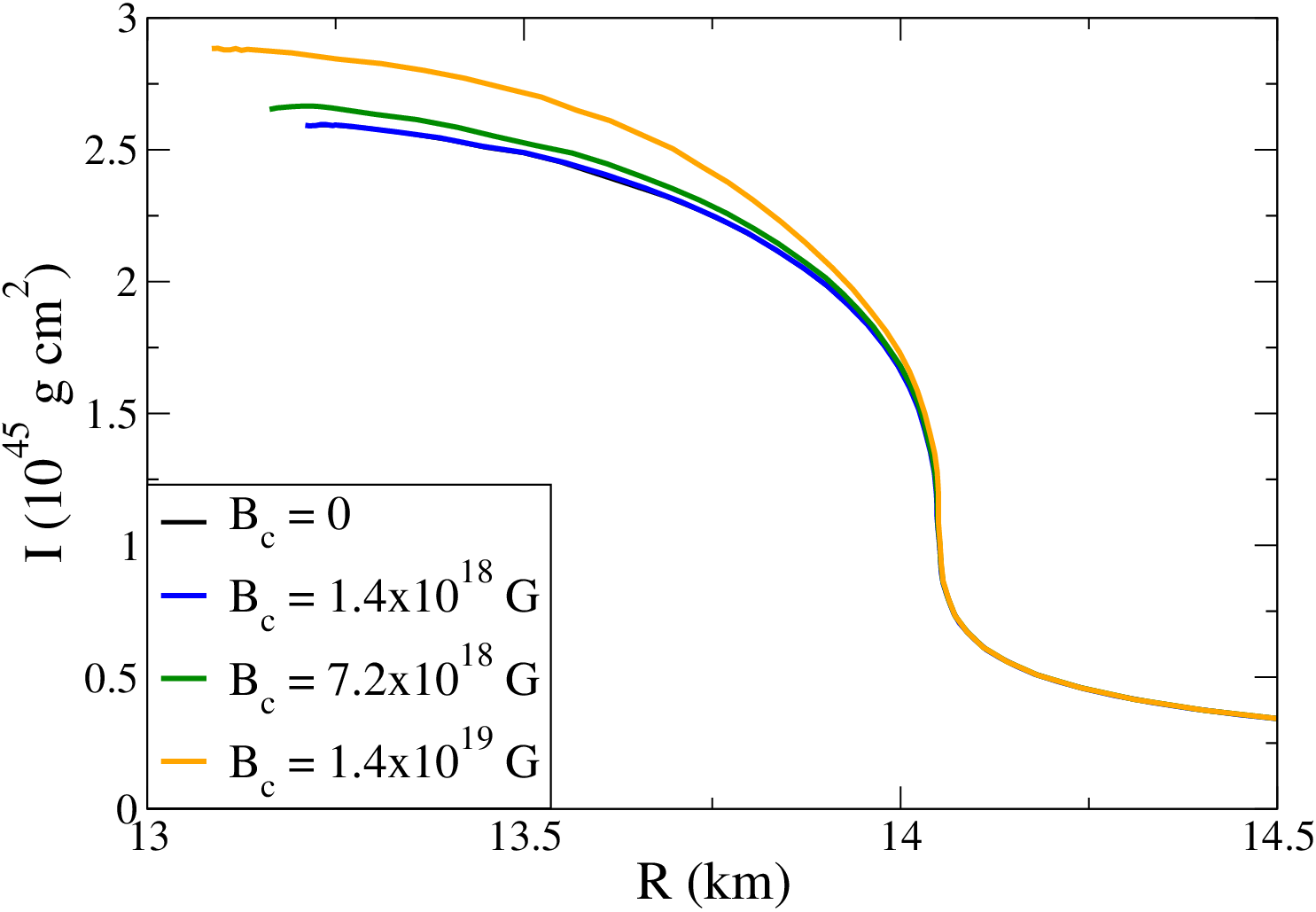}}
\caption{(Color online) Moment of inertia as a function of radius for different
central magnetic fields including AMM.}
\label{I_x_R}
\end{figure}

In Fig.~\ref{I_x_R} we show the moment of inertia as function of radius for
several of the investigated stellar sequences, taking into account the AMM. We
see that for the maximum central magnetic field considered ($1.4\times10^{19}$ G), there is an increase of about 10\% from the unmagnetized case (for the
maximum mass star). For $B_c=7.2\times10^{18}$ G this increase is of about
$\sim$ 2\%. These results indicate that if the magnetic field of the object is
changing, the moment of inertia will be modified, which might affect the spin
evolution of the object, and might allow one to make a connection between AXP
emission and glitches (see discussion in Ref.~\cite{Ryu:2011gq}). For lower
central magnetic fields, the moments of inertia are almost the same as in the
non-magnetized case. 

\subsection{Mass Radius Diagram}
\label{mass radius diagram}

\begin{figure}
\resizebox{0.48\textwidth}{!}{%
\includegraphics{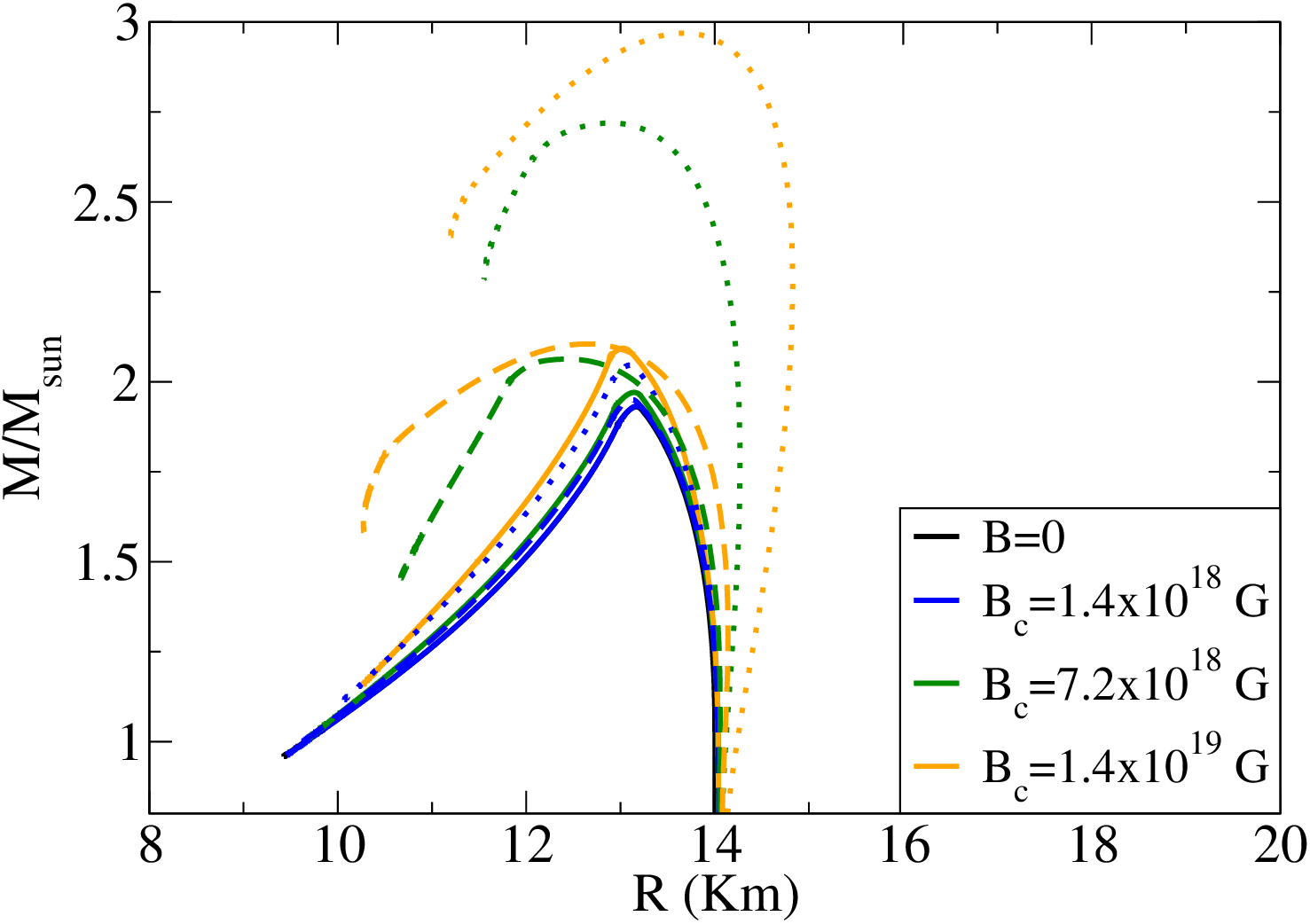}}
\caption{(Color online) Mass-radius diagram shown for different central magnetic
fields including AMM. The dashed and dotted lines have the extra magnetic
pressure term  of ${B^*}^2/24\pi$ and ${B^*}^2/8\pi$, respectively, added to the
total pressure.}
\label{tov}
\end{figure}

As a first approach to the problem, the possible hybrid star masses and radii
are calculated by solving the Tolman-Oppenheimer-Volkoff equations for spherical
isotropic static stars \cite{Tolman:1939jz,Oppenheimer:1939ne} using the EOS of matter. In
Fig.~\ref{tov}, besides our EOS for the core, a separate EOS was used for the crust \cite{Baym:1971pw}. The maximum mass supported
against gravity is higher for higher magnetic fields and even higher when the
AMM is included. It is important to note that in this model pure quark matter
only appears at very high densities, that correspond to the unstable branch of
the mass-radius diagram. Thus, only hadronic and ``mixed" matter appear in the
star. Similar results for stars containing only hadronic and ``mixed matter"
were also found in a calculation using the Brueckner-Hartree-Fock model for the
hadronic phase and the Dyson-Schwinger approach for the quarks
\cite{2011PhRvD..84j5023C}.

So far, the magnetic field energy density and pressure contributions were not considered and only the influence of
the magnetic field on the energy levels of the particles was taken into account.
The problem in including the magnetic pressure is that it has different values
in the directions parallel and perpendicular to the field, as can be seen in the
electromagnetic energy-momentum tensor
\begin{eqnarray}
T^{\mu\nu} = \frac{1}{4\pi}\left( \begin{array}{cccc}
\frac{1}{2}B^{*2} & 0 & 0 & 0 \\
0 & \frac{1}{2}B^{*2} & 0 & 0 \\
0 & 0 & \frac{1}{2}B^{*2} & 0 \\
0 & 0 & 0 & -\frac{1}{2}B^{*2} \end{array} \right),
\end{eqnarray}
where the first term is related to the energy density, the other three terms are related to the
pressure in the $x$, $y$ and $z$ directions and the $4\pi$ comes from the choice of Gaussian natural units. This problem was pointed out in
many papers such as
refs.~\cite{Chaichian:1999gd,Martinez:2003dz,PerezMartinez:2005av,Felipe:2007vb,%
PerezMartinez:2007kw,Ferrer:2010wz,Orsaria:2010xx,Paulucci:2010uj}. In these
papers, there is also a term coming from the magnetic dipole interaction, which
is linear in $B$ and is therefore subleading at extremely high magnetic fields.

A consistent inclusion of the macroscopic magnetic pressure requires a 2D
calculation. As mentioned before, such calculations exist
\cite{Bocquet:1995je,2001ApJ...554..322C}; however, most calculations performed
using realistic EOS's for the macroscopic properties of magnetized stars assume
isotropy and consider the pure magnetic pressure term to be either positive or negative in all
directions. It was pointed out by
refs.~\cite{Bandyopadhyay:1997kh,Sinha:2010fm} that considering the magnetic pressure to be negative in all
directions constrains the maximum values that can be used for the magnetic field
to lower values. In our case,
$B_c=7.2\times10^{18}$ G would already cause an instability at very high
densities. A third option, suggested by Ref.~\cite{Bednarek:2002hb} uses, as a
monopole approximation of the energy-momentum tensor, the average between the
three pressures and adds $+B^2/24\pi$ for the magnetic pressure in all
directions. We show the difference caused by either choosing the pure magnetic pressure to be
positive in all directions or using 
the average pressure in the mass-radius diagram. Evidently, the first case is
unphysical (for extremilily strong magnetic fields) since the magnetic pressure dominates over any other contribution. We
speculate that the correct value would be close to the second option that uses
the average of the pressures in different directions, but any more precise
statement requires the use of axisymmetric equations, as was already pointed by
Ref.~\cite{Paulucci:2010uj}. Work along this line using our hybrid matter EOS is
ongoing and will allow for a more exact estimate of the increase of the star
mass based on the magnitude of the magnetic field of the magnetar.

\subsection{Thermal Evolution}
\label{thermal evolution}

We now turn our attention to the thermal evolution of hybrid stars, whose
microscopic composition is given by the model described in this paper. The
cooling of compact stars is given by the thermal balance and thermal energy
transport equation ($G = c = 1$) \cite{Weber}
\begin{eqnarray}
  \frac{ \partial (l e^{2\phi})}{\partial m}& =   &-\frac{1}{\rho \sqrt{1 -
2m/r}} \left( \epsilon_\nu e^{2\phi} + c_v \frac{\partial (T e^\phi) }{\partial
t} \right) \, , 
  \label{coeq1}  \\
  \frac{\partial (T e^\phi)}{\partial m} &=& - 
  \frac{(l e^{\phi})}{16 \pi^2 r^4 \kappa \rho \sqrt{1 - 2m/r}} 
  \label{coeq2} 
  \, .
\end{eqnarray}
In Eqs.~(\ref{coeq1})$-$(\ref{coeq2}) the structure of the star is given by the
variables  $r$, $\rho(r)$, $e^{\phi}$ and $m(r)$, that represent the radial
distance, the energy density, the metric function, and the stellar mass,
respectively. The thermal variables are given by the temperature $T(r,t)$,
luminosity $l(r,t)$, neutrino emissivity $\epsilon_\nu(r,T)$, thermal
conductivity $\kappa(r,T)$, and specific heat $c_v(r,T)$. The boundary
conditions of (\ref{coeq1}) and (\ref{coeq2}) are determined by the luminosity
at the stellar center and at the surface. The luminosity vanishes at the stellar
center since there is no heat flux there.  The surface temperature (luminosity)
is defined by the crust temperature and the properties of the stellar surface
(surface gravity, magnetic field and etc.)
\cite{Gudmundsson1982,Gudmundsson1983,Page2006,Blaschke:1999qx}.

For the
hadronic phase, we have considered the following neutrino emission processes:
direct Urca, modified Urca and bremsstrahlung processes; whereas 
for the quark phase, the processes taken into account are the quark direct Urca,
quark modified Urca, and quark bremsstrahlung processes. Details about the
emissivities of such processes can be found in
refs.~\cite{Yakovlev2001a,IWAMOTO1982}. The specific heat of the hadrons is the
usual specific heat of fermions, as described in Ref.~\cite{Page:2004fy}. As for
the quarks, we use the expression for the specific heat calculated in
Ref.~\cite{IWAMOTO1982}. The thermal conductivities for the hadronic and quark
matter were calculated, respectively in refs.~\cite{Flowers1981,Haensel1991},
and in this work we follow the results presented in these references. 

We have calculated the cooling for magnetized hybrid stars (with anomalous
magnetic moment of the hadrons included) for three different masses: 1.1, 1.4
and 1.93 $M_\odot$. For each hybrid star mass we considered three values for the central
magnetic field ($1.4\times10^{18},7.2\times10^{18},1.4\times10^{19}$ G). In addition to
that we have also considered the effect of including the pure magnetic field contribution in the EoS.

\begin{figure}
\resizebox{0.48\textwidth}{!}{%
\includegraphics{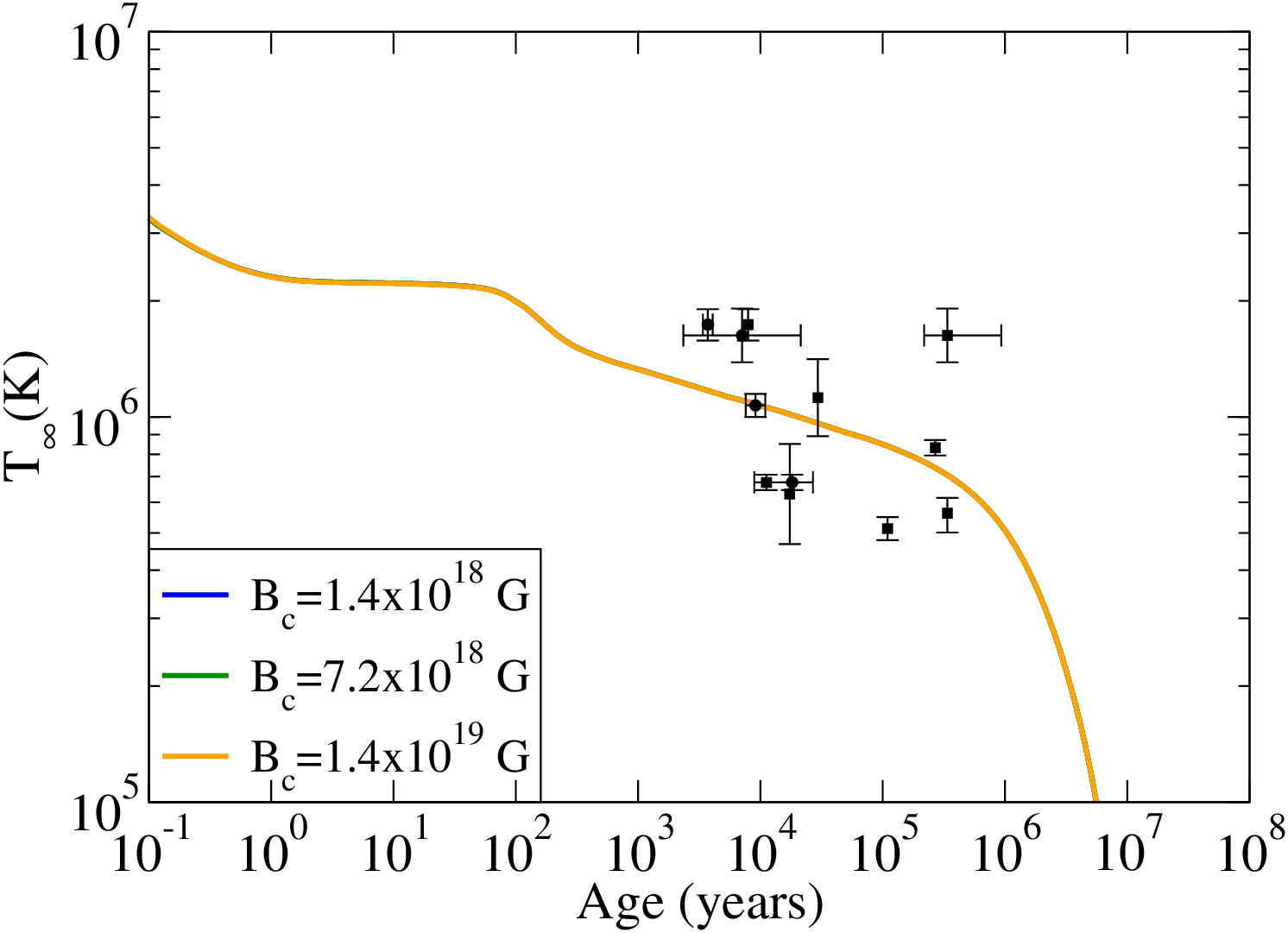}}
\caption{(Color online) Cooling curves for a 1.1 $M\odot$ mass star in the presence of different central magnetic fields including AMM. $T_\infty$ denotes the redshifted temperature observed at infinity. The observed data consists of circles denoting spin-down ages and squares denoting kinematic ages \cite{Page:2004fy,Page:2009fu}. The dashed and dotted lines include the extra magnetic pressure term of ${B^*}^2/24\pi$ and ${B^*}^2/8\pi$, respectively. All curves overlap.}
\label{cooling11}
\end{figure}

The cooling of 1.1 $M_\odot$ stars is shown in Fig.~\ref{cooling11}. One can see
that in this case the star exhibits a slow cooling, which agrees relatively well
with the observed data. Furthermore, we can also conclude that in this case the
magnetic field has no substantial effect on the thermal evolution of the object,
and neither does the inclusion of the pure magnetic contribution in the EoS.

\begin{figure}
\resizebox{0.48\textwidth}{!}{%
\includegraphics{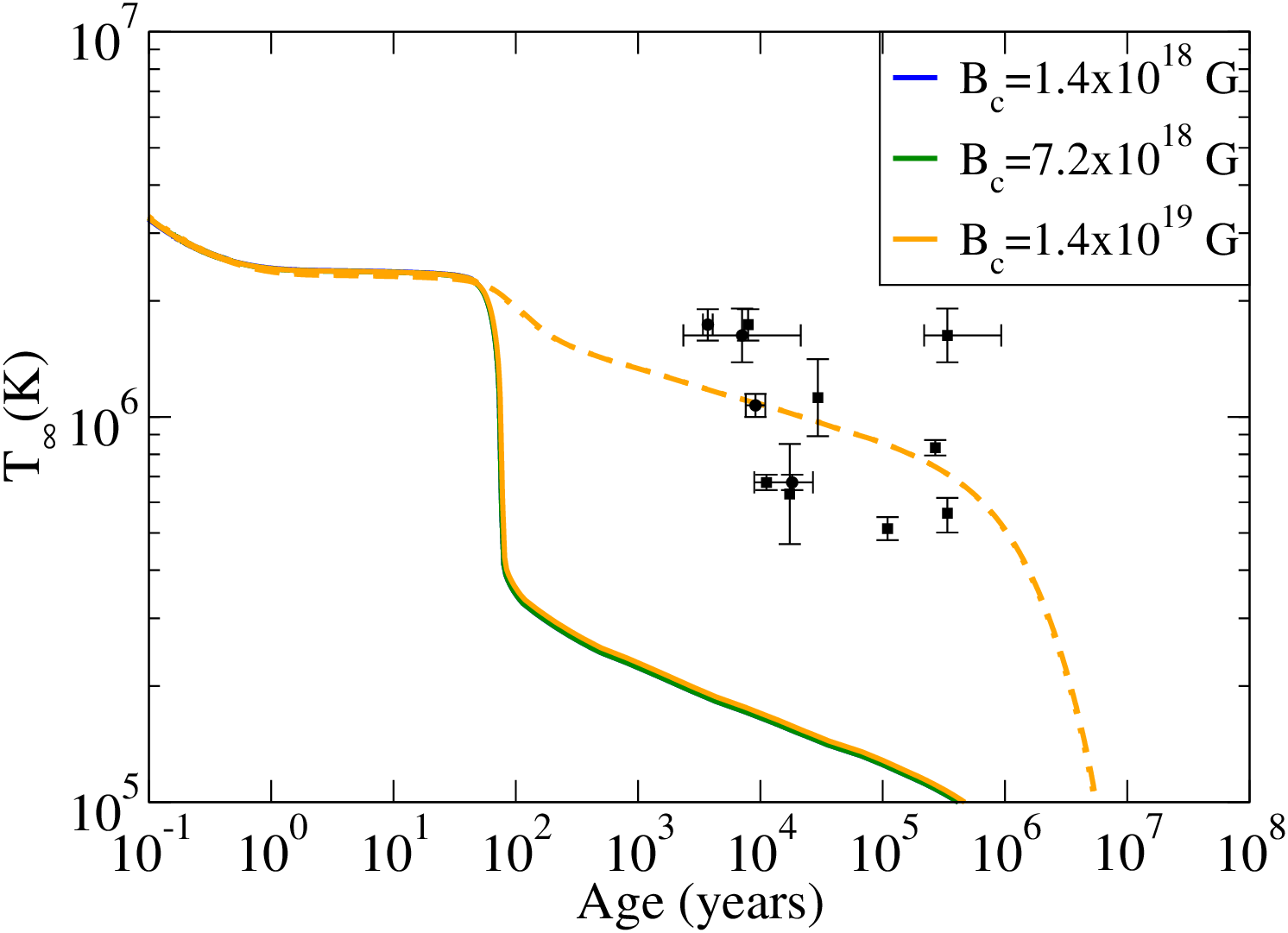}}
\caption{(Color online) Cooling curves for a 1.4 $M\odot$ mass star. Otherwise same notation as Fig.~\ref{cooling11}. The dashed and dotted lines curves overlap.}
\label{cooling14}
\end{figure}

\begin{figure}
\resizebox{0.48\textwidth}{!}{%
\includegraphics{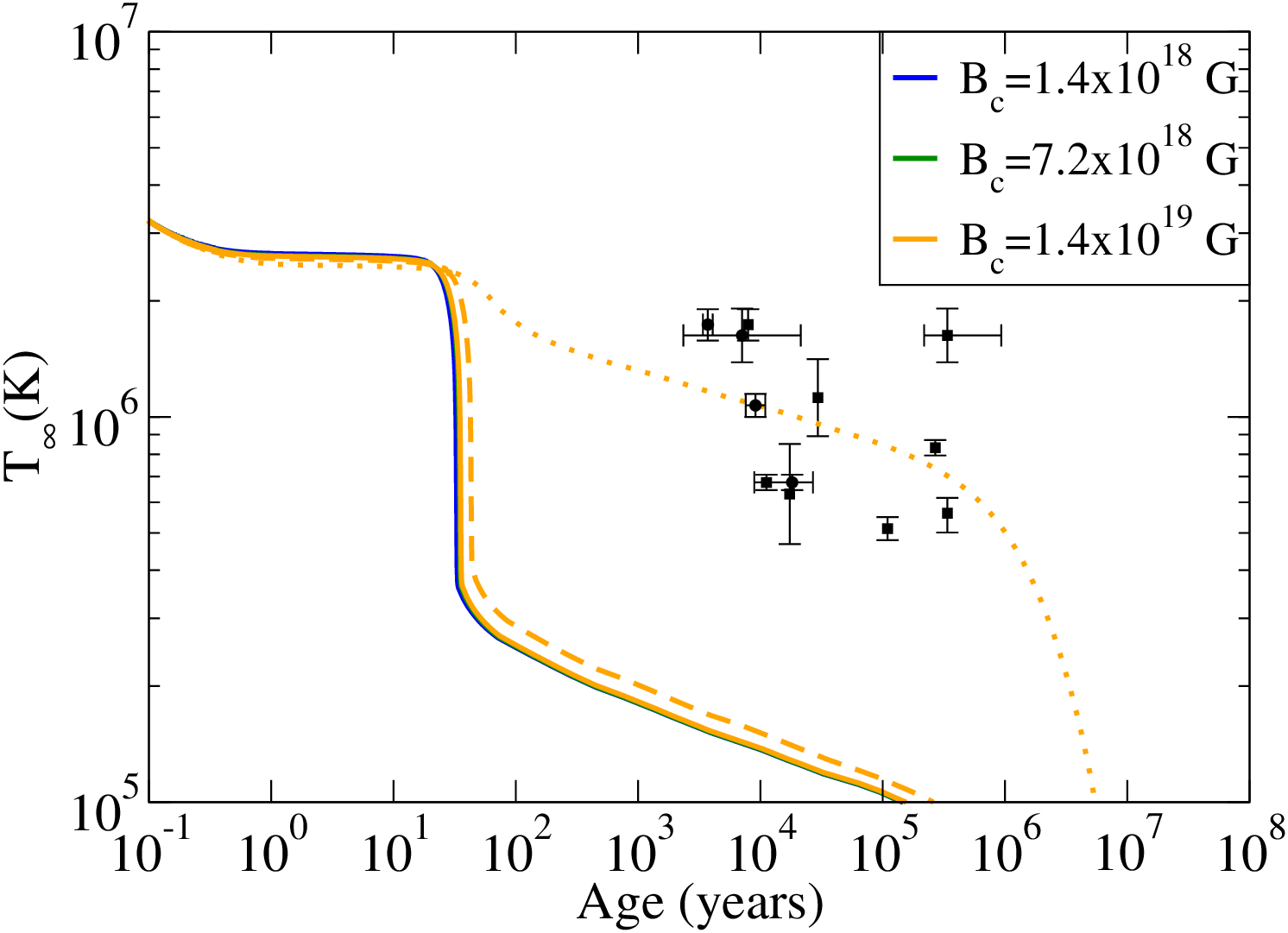}}
\caption{(Color online) Cooling curves for a 1.93 $M\odot$ mass star. Otherwise same notation as Fig.~\ref{cooling11}.}
\label{cooling193}
\end{figure}

The situation is similar for stars with higher masses, as can be seen by the full lines in 
Figs.~\ref{cooling14} and \ref{cooling193}, which show the thermal evolution of
stars with 1.4 and 1.93 $M_\odot$, respectively. We see that in this case the
modifications introduced by the magnetic field in the composition are not enough
to alter the cooling significantly (as was the case for lower mass stars). This picture
changes, however, if one introduces the effect of the magnetic pressure. As
shown by the dotted lines in Figs.~\ref{cooling14} and \ref{cooling193}, the extra $B^{*2}/8\pi$
leads to the slow cooling of a star that would otherwise exhibit fast cooling.
This stems from the fact that adding an extra source of pressure stiffens the
EoS. Hence, stars with higher masses possess smaller central densities and,
therefore, smaller proton fractions. The small proton fraction will hinder the
otherwise present direct Urca process, thus leading to a slow cooling.
 The
results for the inclusion of the magnetic pressure of $B^{*2}/24\pi$ (dashed lines) are
qualitatively the same. In 
this case, however, the stiffening of the EoS is more mild, and thus the effect
is only appreciable for moderate masses, as seen in Fig.~\ref{cooling14}, where
the thermal evolution is the same as for
the magnetic pressure of $B^{*2}/8\pi$. For the 1.93 $M_\odot$ star we see that
the stiffening of the EoS is not enough to hinder the direct Urca process
completely, and cooling of hybrid stars with magnetic pressure of
$B^{*2}/24\pi$ is only slightly slower.

 Note that in Figs~ 9, 10 and 11 we have
assumed an effective magnetic field that changes with density within the star
but does not change in time. In a realistic scenario it is expected that the
magnetic field decreases as a function of time as explained in
Ref.~\cite{Aguilera:2007dy,Aguilera:2007xk,Xie:2011bd}. In these references the
authors also conclude that 
the inclusion of the magnetic field directly in the cooling simulation
reproduces stars that remain warmer for a longer time, in better agreement with
observation. On the other hand, such studies, do not include the effect of
magnetic fields in the EOS, and therefore are complemented by our study.

\subsection{Conclusions}
\label{conclusions}

To model magnetars, we use an effective model that includes hadronic and quarks
degrees of freedom. As the density increases, the order parameters signal the
deconfinement and chiral symmetry phase transitions, which tend to take
place closer to the the center of the star for higher magnetic fields. The
stiffness of the equation of state (EOS), and the consequent star masses, depend on the chosen
central magnetic field. It is clear that a higher $B_c$ allows for more massive
stars, but the quantitative analysis of how more massive the star might be
requires the use of a 2D solution of Einstein's equations which takes into
account the breaking of spherical symmetry by the magnetic field. This was shown by the dramatic change in the star masses when different assumptions were used for the inclusion of the pure magnetic field contribution to the EOS. As a consequence, only the lower magnetic field configuration with $B_c=1.4\times 10^{18}$ gives reliable results, as it has a pressure anisotropy $(p_z-p_{x,y})/(p_z+p_{x,y})$ of $\sim 20\%$. Having a 2D formalism
will also allow to calculate the exact hydrodynamic limit for the magnetic field
for our model. Efforts along these lines, using our EOS are currently taking
place. In addition to the effect on the mass-radius relationship, we have shown
that the presence of high magnetic fields can result in peaks in the adiabatic
index and can modify the star's moment of inertia.

We introduce an expression for the effective magnetic field $B^*$ that increases
with baryon chemical potential, and therefore avoids a discontinuity in the
phase transition region. This allows $B^*$ to increase smoothly from a surface
value up to a chosen central one ($B_c$). Furthermore, our investigation
indicates that the composition changes introduced by the magnetic field are
not high enough to alter significantly the thermal evolution of hybrid stars. We have found,
however, that by including the magnetic pressure ($B^{*2}/8\pi$ or
$B^{*2}/24\pi$), the consequent stiffening of the EoS allows for stars with
higher masses (1.4~--~1.93 $M_\odot$) to exhibit slower cooling, which is in
better agreement with the observed data. This is an important result, which
indicates the importance of correctly introducing the macroscopic pressure
 into the structure of the star. We are cautious in interpreting these
results, since the appropriate treatment of both the structure, and the thermal
evolution of stars with  such high magnetic fields requires a full
two-dimensional study \cite{Negreiros:2012aw}. We will save this investigation
for a future work, as it seems clear to us that the effect of the magnetic
pressure on the structure and thermal evolution are substantial. Although
calculations which address this problem using model EOS's have
already been performed, we believe that a well-motivated microscopic EOS that also includes the effects of the star's magnetic field is important
for fully understanding the behavior of magnetars. 

Subsequently, we would like to expand our study to finite temperature, in order
to have a complete phase diagram (temperature as a function of baryon chemical
potential) for high magnetic fields. A phase diagram for $B=0$ was already
constructed using the extended non-linear realization of the SU(3) sigma model
in Ref.~\cite{Dexheimer:2009hi}. Such a complete analysis is important, since it
connects the physics at high temperature and small densities such as those
created in heavy ion collisions, with the physics inside neutron stars. A
finite-$B$ phase diagram was already constructed in
Ref.~\cite{Chakrabarty:1996te} but we believe that we can give further insight
into the matter with our model, since it also includes chiral symmetry
restoration and allows for a smooth chiral and deconfinement crossover
transition at small densities. It would also be interesting to study different
parametrizations of the model which would give a smooth crossover in the low
temperature regime and study the influence of the 
magnetic field in that case.

\section*{Acknowledgments}
We are grateful to Constanca Providencia, Milva Orsaria, Fridolin Weber, Debora
Menezes and Aurora Perez Martinez for the fruitful discussions on the subject of
strong magnetic fields. R. N. acknowledges financial support from the LOEWE
program HIC for FAIR. V. D. acknowledges financial support from CNPq-Brazil.

%
 \bibliographystyle{epj}
 \bibliography{template_}
%
%
%

\end{document}